\documentclass{article}

\usepackage{spconf,amsmath,graphicx}

\usepackage[natbib, bibencoding=utf8, citestyle=numeric-comp, bibstyle=ieee, maxbibnames=999, maxcitenames=2, mincitenames=1, sortcites]{biblatex}
\bibliography{refs.bib}

\usepackage{tabularx}
\usepackage{tabulary}
\usepackage{makecell}

\usepackage{multirow}
\usepackage{subcaption}
\usepackage{booktabs}
\usepackage{threeparttable}

\usepackage{hyperref}

\usepackage[capitalise, nameinlink, noabbrev]{cleveref}

\usepackage{todonotes}

\usepackage{caption}
\usepackage{verbatim}
\usepackage{listings}

\usepackage{svg}

\usepackage{balance}

\usepackage[inline]{trackchanges}

\usepackage[acronym, shortcuts, nohypertypes={acronym}]{glossaries}
\newacronym{ACC}{ACC}{accuracy}
\newacronym{AST}{AST}{audio spectrogram transformer}
\newacronym{BE}{BE}{biodiversity exploratories}
\newacronym{CNN}{CNN}{convolutional neural network}
\newacronym{DL}{DL}{deep learning}
\newacronym{MU}{$\mu$}{mean $\mu$}
\newacronym{ML}{ML}{machine learning}
\newacronym{NLP}{NLP}{natural language processing}
\newacronym{RHO}{$\rho$}{Spearman's $\rho$}
\newacronym{SIGMA}{$\sigma$}{standard deviation $\sigma$}
\newacronym{SGD}{SGD}{stochastic gradient descent}
\newacronym{UAR}{UAR}{unweighted average recall}
\newacronym{ZSL}{ZSL}{zero-shot learning}

\def\x{{\mathbf x}}

\title{Exploring meta information for audio-based\\ zero-shot bird classification}
\name{
Alexander Gebhard\textsuperscript{1,2},
Andreas Triantafyllopoulos\textsuperscript{1,2},
Teresa Bez\textsuperscript{2},\\
\textit{Lukas Christ\textsuperscript{2}},
\textit{Alexander Kathan\textsuperscript{2}},
\textit{Bj\"orn W. Schuller\textsuperscript{1,2,3}}
}
\address{
  \textsuperscript{1}CHI -- Chair of Health Informatics, MRI, Technical University of Munich, Germany\\
  \textsuperscript{2}Chair of Embedded Intelligence for Health Care and Wellbeing, University of Augsburg, Germany\\
  \textsuperscript{3}GLAM -- Group on Language, Audio, \& Music, Imperial College London, UK
}

\begin{document}
%
\maketitle
\begin{abstract}
Advances in passive acoustic monitoring and machine learning have led to the procurement of vast datasets for computational bioacoustic research. 
Nevertheless, data scarcity is still an issue for rare and underrepresented species. 
This study investigates how meta-information can improve zero-shot audio classification, utilising bird species as an example case study due to the availability of rich and diverse metadata.
We investigate three different sources of metadata: textual bird sound descriptions encoded via \textsc{(S)Bert}, functional traits (\textsc{Avonet}), and bird life-history (\textsc{BLH}) characteristics. 
As audio features, we extract audio spectrogram transformer (AST) embeddings and project them to the dimension of the auxiliary information by adopting a single linear layer.
Then, we employ the dot product as compatibility function and a standard zero-shot learning ranking hinge loss to determine the correct class. 
The best results are achieved by concatenating the \textsc{Avonet} and \textsc{BLH} features attaining a mean unweighted F1-score of \textbf{$.233$} over five different test sets with $8$ to $10$ classes.
\end{abstract}

\section{Introduction}
\label{sec:intro}
The field of computational bioacoustics has recently witnessed tremendous growth thanks to rapid technological progress, and in particular by exploiting the recent advances in machine learning~\citep{Stowell22-CBW}. 
The availability and affordability of good hardware, such as microphones or storage devices, vastly expands the capabilities of bioacoustic monitoring \citep{Stowell22-CBW}. 
It is now possible to continuously capture audio at multiple and large areas at the same time, which leads to an enormous amount of audio data that needs to be processed \citep{Kahl21-BAD,Stowell22-CBW}. 
As a result, experts do not have enough time and resources to analyse or annotate the data on their own without automated processes, which makes the use of computational methods imperative.
These methods, once trained on sufficient amounts of data, can be used to speed up the detection of species through their vocalisations.
However, there is a plethora of rare or threatened species for which there may not be enough data to train an initial model \citep{Borgelt22-MTH}; yet, they in particular are most interesting from a biodiversity perspective, making their successful detection a crucial aspect of monitoring and conservation efforts. 
This is where \ac{ZSL} could be applied to annotate audio samples without any previous labelling efforts, relying only on external meta information. 
This auxiliary information can be textual annotations of sound classes or events \citep{Elizalde23-CLA,Xie21-ZAC}, coming from other modalities like images \citep{Dogan22-ZAC}, or even from multiple modalities at the same time \citep{Guzhov22-AEC}.


Important advancements in \ac{ZSL} have been primarily achieved in the computer vision domain, and rely on semantic information such as text data \citep{Socher13-ZLT,Frome13-DAD,Radford21-LTV}.
After the initial success in the visual domain, the computer audition community also adopted and refined \ac{ZSL} approaches for their tasks \citep{Xie21-ZAC,Dogan22-ZAC}. Among the recent breakthroughs are adaptations of the \textsc{CLIP} method from computer vision, such as \textsc{AudioClip} \citep{Guzhov22-AEC}, \textsc{Wav2CLIP} \citep{Wu22-WLR}, or \textsc{CLAP} \citep{Elizalde23-CLA}.

The visual recognition of avian species has become a standardised benchmark for \ac{ZSL}, owing to the importance of the problem and the availability of suitable metadata: binary description attributes~\citep{Wah11-TCB}, textual descriptions~\citep{Reed16-LDR}, field-guide visualisations~\citep{Rodriguez22-ZBS}, and even DNA data~\citep{Badirli21-FZL} can all serve as mediating attributes for \ac{ZSL}.
Yet, while the visual recognition of birds is a vital aspect of biodiversity research, the more pressing issue of zero-shot auditory recognition has not received as much attention, despite the fact that audio offers improved monitoring capabilities for birds in the high-occlusion conditions of their natural habitats.
While considerable efforts have been expended in closed-set \citep{Kahl20-OOB} and few-shot recognition~\citep{Wu23-FCL,You23-TBS}, and some recent interest in open-set recognition~\citep{Xie23-COS}, we have found no previous works investigating the application of \ac{ZSL} on audio-based recognition of birds.
This is a gap we attempt to mitigate in the present contribution.

Specifically, we aim to identify the most promising form of metadata that can serve as mediating variables for \ac{ZSL}.
Our starting point is a dataset of $95$ European bird species assembled from \textsc{Xeno-Canto}.
These particular species have been selected based on a recent survey from \citet{Jung12-BSD} on the presence of avian species in the areas monitored by an ongoing, large-scale biodiversity programme, the \ac{BE}\footnote{\label{fn:BE-exploratories}\url{https://www.biodiversity-exploratories.de/en/}}.
We explore the following alternative forms of metadata: a) vocalisation descriptions extracted from a standardised field-guide, b) aggregated morphological attributes, and c) life-history traits.
For our modelling, we draw on recent works on audio-based \ac{ZSL} and rely primarily on a simple \ac{ZSL} model~\citep{Xie21-ZAC} -- our goal is to establish a baseline and not go after state-of-the-art results. The corresponding code can be found on github\footnote{\url{https://github.com/ATriantafyllopoulos/audiocub-zsl}}.


\section{Dataset}
\label{sec:dataset}
For our experiments, we utilise audio data and meta information from $95$ European bird species. The 95 birds were chosen based on a previous survey by \citet{Jung12-BSD} from the \ac{BE}
\footref{fn:BE-exploratories}, since our ultimate goal will be to deploy our \ac{ZSL} model to automatically annotate the audio data of this project.
The audio data was collected from \textsc{xeno-canto} while the auxiliary information was taken from three different sources to investigate their influence on the model performance. 
The audio data from \textsc{xeno-canto} are in MP3 format and comprise roughly $725$ hours.

The first type of metadata was taken from the standard Princeton Field Guide by \citet{Svensson10-BOE}, which contains descriptions of bird sounds for the 95 species. That is, the sound of a bird species is described in a textual manner w.\,r.\,t.\ specific patterns and peculiarities, while relying heavily on onomatopoeia. 
The following quote, belonging to the species \textit{phoenicurus ochruros}, gives an impression of these descriptions:
\begin{quote}
\scriptsize{Call a straight, slightly sharp whistle, `vist', often repeated impatiently.
When highly agitated, a discreet clicking is added, `vist, tk-tk-tk'. Song
loud, frequently given at first light from high perch, usually consists of
four parts: starts with a few whistles and a rattling repetition of same note,
followed by a pause c.\ 2 sec.\ long, then a peculiar crackling sound (not very
far-carrying), after which the verse terminates with some brief whistled notes,
e.\,g. `si-sr\"u TILL-ILL-ILL-ILL-ILL....... (krschkrschkrsch) SR\"Usvisvi'; the
sequence of the four components may sometimes be switched around.}
\end{quote}

The \textsc{Avonet} dataset~\cite{Tobias22-AME}, comprising ecological parameters, continuous morphological traits, and information on range and location, was the second source of auxiliary data.
\textsc{Avonet} includes the following parameters: beak and wing measurements; tarsus and tail length; kipps distance; mass; habitat; habitat density; primary lifestyle; etc.
The dataset was collected by \citet{Tobias22-AME} to enable theoretical tests as well as the investigation of evolutionary biology, ecology, and the functioning of biodiversity. The data was gathered from literature, fieldwork, and various museum collections. 

The dataset from \citet{Storchova18-DFL}, comprising morphological, reproductive, behavioural, dietary, and habitat preference characteristics, was the third source. For the rest of this work, the third set of metadata will be simply denoted as \textsc{BLH}.
The \textsc{BLH} 
features include the following traits: average egg length, weight, and mass; type of nest; average length, wingspan, tailspan, bill size, tarsus, and weight separately for male and female members of each species; incubation period; age of first breeding; etc. Binary indicators for nesting preferences and feeding habits are also included.
These statistics are collected from a sample of European birds obtained primarily from the Birds of the Western Palearctic handbook to provide an open access dataset for ``research investigating large-scale patterns in European avifauna'' \citep{Storchova18-DFL}. 


\begin{figure}
    \centering
    \includegraphics[width=\linewidth]{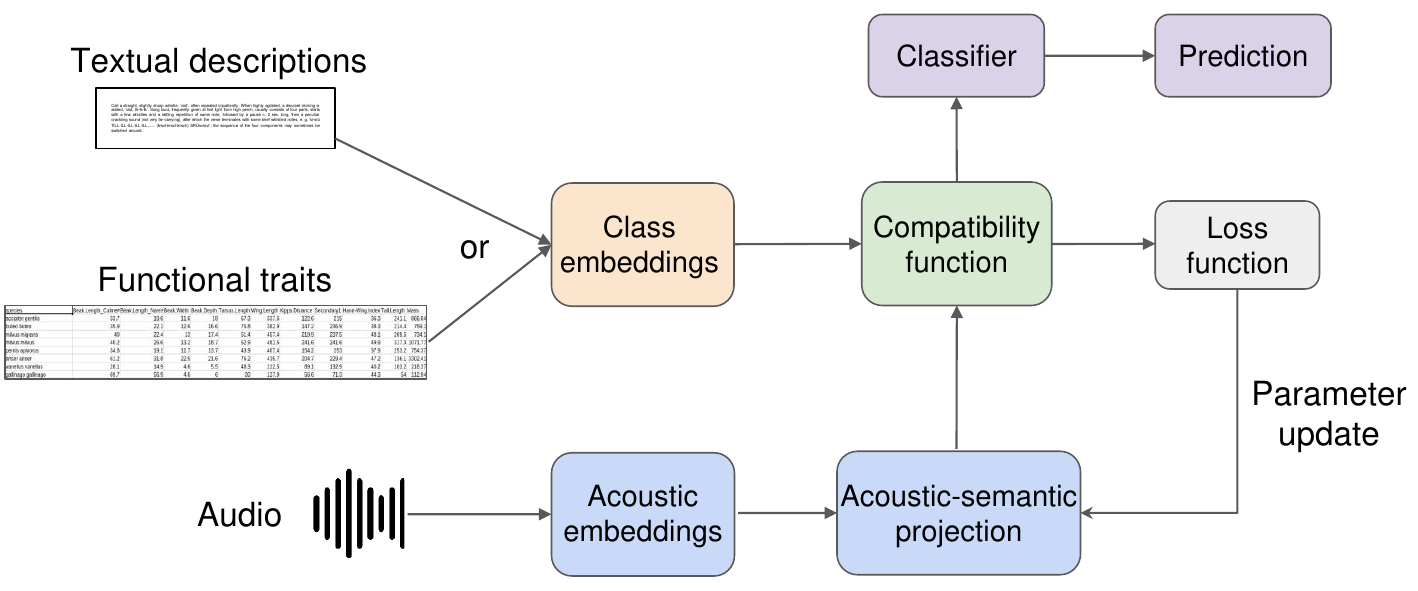}
    \caption{The utilised \ac{ZSL} pipeline, based on \citep{Xie21-ZAC}.}
    \label{fig:pipeline}
\end{figure}


\section{Methodology}
\label{sec:methodology}

This section outlines the utilised features, the zero-shot classification approach, and the experimental setup of our study. 

\subsection{Features}
\label{ssec:features}
We employ \ac{AST} embeddings as audio features which are extracted by a state-of-the-art \ac{AST} model\footnote{\url{https://huggingface.co/docs/transformers/model_doc/audio-spectrogram-transformer}} \citep{Gong21-AAS}. Before extracting the embeddings, we resample the audio files to $16$\,kHz. For each audio file, we get a 2D array and average it over time to obtain a 1D vector with a dimension of $768$. 


In order to acquire meaningful representations of the textual bird sound descriptions, we adopt two pre-trained Transformer-based language models. First, we utilise \textsc{Bert} \cite{devlin-etal-2019-bert} in its base variant (110M parameters)\footnote{\url{https://huggingface.co/bert-base-uncased}}.~\citet{reimers-2019-sentence-bert} show that this type of model may not be optimal for Semantic Textual Similarity tasks and thus propose \textsc{Sentence-Bert} (\textsc{SBert}), specifically optimised to compute sentence and paragraph embeddings that reflect semantic similarity. From the set of pre-trained \textsc{SBert} models provided in their library\footnote{\url{https://www.sbert.net}}, we opt for the paraphrase-multilingual-mpnet-base-v2 model\footnote{\url{https://huggingface.co/sentence-transformers/paraphrase-multilingual-mpnet-base-v2}}.
For each bird, we extract both \textsc{Bert} and \textsc{SBert} feature vectors of size $768$ by feeding the bird's 
entire sound description into the respective model. From the \textsc{Bert} model, the averaged embeddings of the final layer are taken as the text representation, while the \textsc{SBert} model's embedding is obtained via the provided pooling method.

\begin{figure}
    \centering
    \includegraphics[width=.8\columnwidth]{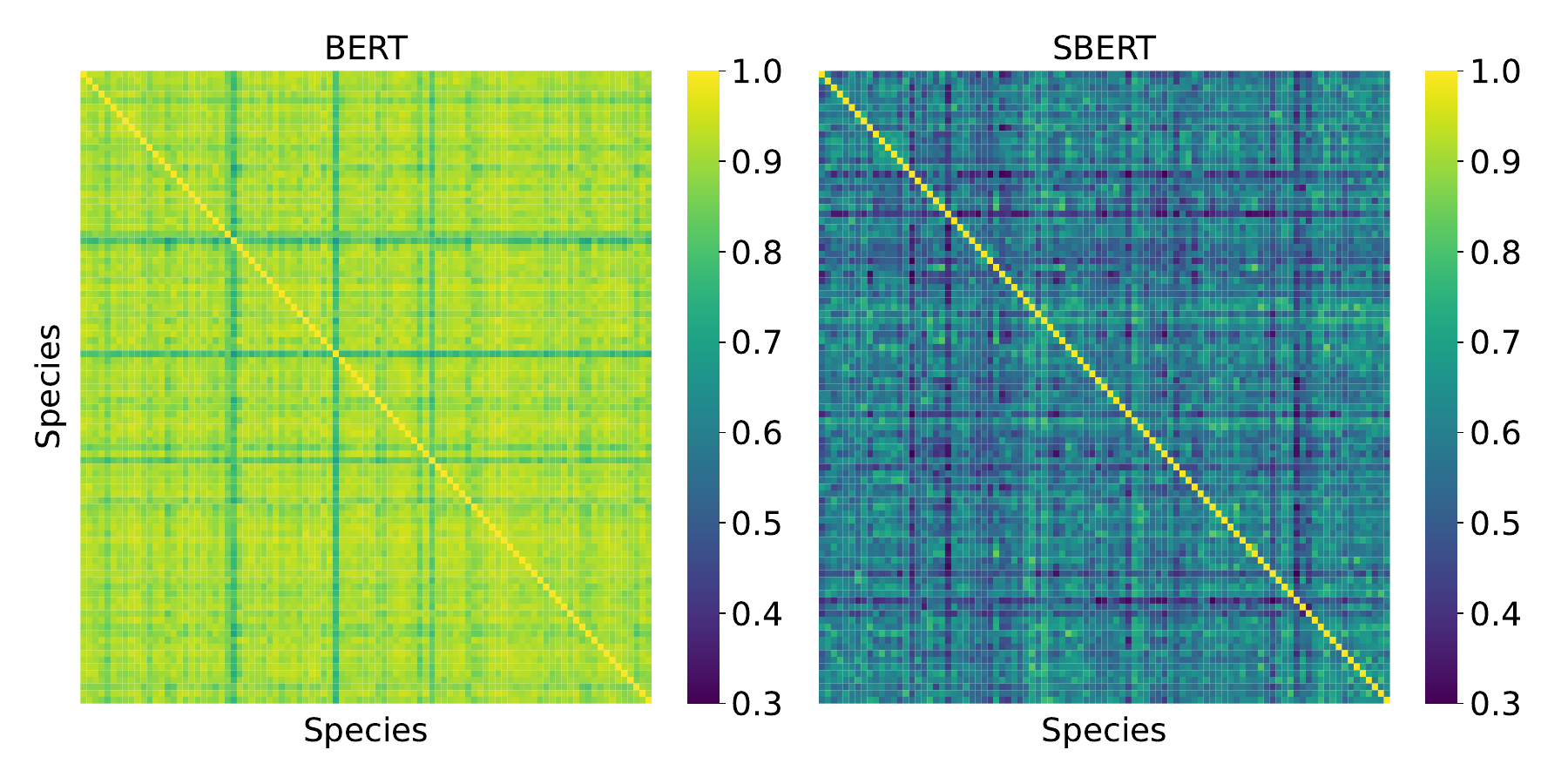}
    \caption{The pairwise cosine similarities between the $95$ bird species for the \textsc{Bert} embeddings (left) and the \textsc{SBert} embeddings (right) visualised as heat maps. The \textsc{SBert} embeddings show a stronger difference among the birds.}
    \label{fig:heatmaps}
\end{figure}

In this context, we analyse the pairwise cosine similarities among the bird species' embeddings to see which of the two textual embedding methods creates stronger distinctions. That is, for each bird species we compute the cosine similarity between the \textsc{(S)Bert} embeddings of this species and every other species, yielding a matrix of cosine similarities w.\,r.\,t.\ those embeddings. These matrices are visualised as heat maps in \Cref{fig:heatmaps} and suggest that \textsc{SBert} yields more distinguishable representations for the species.
Therefore, we expect \textsc{SBert} to achieve a better model performance than \textsc{Bert}. The corresponding results are reported and discussed in \Cref{sec:results}.

Regarding the \textsc{Avonet} features, we keep only information unrelated to the species name or family and also discard the geographical attributes, such as latitude and longitude, since they do not uniquely characterise each species. We also omit the species or family related information for the \textsc{BLH} dataset. 
Additionally, all attributes that have more than $10$ NaN values are dropped for both feature sets, ending up with $23$ and $77$ features, respectively. All remaining NaN values are set to $0$. Finally, each feature being a string is encoded to a numerical representation with a common label encoder. Before these features are then fed to the model, 
each of them is scaled to the range $[0,1]$ via min/max normalisation.

\subsection{Zero-Shot Classification}
\label{ssec:zero-shot}
The \ac{ZSL} procedure we employ in this study builds on the approach presented in \citep{Xie21-ZAC}. They rely on previous work from \citet{Weston10-LSI} and \citet{Akata15-LFI} and apply a compatibility function to an acoustic-semantic projection in order to classify a sound class. Their training procedure involves a ranking hinge loss that exploits the compatibility of the projections. The sound class with the highest compatibility is considered the correct class.

Out of the features from \Cref{ssec:features}, we employ the \ac{AST} representations as our acoustic embeddings $A(x)$ and the \textsc{(S)Bert} embeddings, the \textsc{Avonet}, as well as the \textsc{BLH} features as class embeddings $C(y)$. A high-level overview of our pipeline is visualised in \Cref{fig:pipeline}. In order to project the acoustic to the class embeddings, we utilise a single linear layer which has as many neurons as the size of the respective class embeddings, as done by \citet{Xie21-ZAC}: 

\begin{equation}
    P(A(x)) = W^{T}A(x)
    \label{equ:projection}
\end{equation}

\noindent In order to check the compatibility between the projected acoustic and the class embeddings, we adopt the dot product. 

\begin{equation}
    F(P(A(x)),C(y)) = P(A(x))^{T}C(y)
\end{equation}

\noindent This compatibility function is later exploited in the ranking loss function which is the same as in \citep{Xie21-ZAC,Weston10-LSI}. The goal is that the highest ranked class embeddings best describe the audio sample. 
Thus, after computing the compatibility, the ranks $r_{y_n}$ for each batch element are determined and the corresponding penalties $\rho(r_{y_n})$ are calculated as

\begin{equation}
    \rho(r_{y_n}) = \sum_{i=1}^{r_{y_n}} \frac{1}{i}
\end{equation}

\noindent with $\rho(0) = 0$.
Then, a version of the hinge loss $h$ is computed by employing the compatibility function as in \citep{Xie21-ZAC}, with $\Delta(y_n, y) = 0$ if $y_n = y$ and 1 otherwise:

\begin{align}
    h(x_n, y_n, y) = \Delta(y_n, y) + F(P(A(x_n)),C(y)) \nonumber \\- F(P(A(x_n)),C(y_n))
\end{align}

\noindent Following the weighted approximate-rank pairwise
(WARP) loss from \citet{Weston10-LSI}, our final ranking hinge loss is

\begin{equation}
    \frac{1}{N} \sum_{n=1}^{N} \frac{\rho(r_{y_n})}{r_{y_n}} \sum_{y} \max({0, h(x_n, y_n, y)}), 
\end{equation}

\noindent where $0/0 = 0$ if $r_{y_n} = 0$. 

\subsection{Experimental Setup}
\label{ssec:experimental-setup}

In order to obtain robust results, we create five different splits, each of which consists of a training ($\sim$$80$\%), development ($\sim$$10$\%), and test ($\sim$$10$\%) set. We make sure that each dev and test set comprises different species than the other four dev/test sets, i.\,e., they are disjoint. 
Our experiments investigate how well the \ac{ZSL} approach described in \Cref{ssec:zero-shot} performs with the three meta information sources described in \Cref{sec:dataset}. For this purpose, we employ the \textsc{Bert}, \textsc{SBert}, \textsc{Avonet}, and \textsc{BLH} features introduced in \Cref{ssec:features}. The experiments are conducted on the five splits and the mean of the performance on the dev/test sets are reported.
We train for $30$ epochs and utilise a \ac{SGD} optimiser with a learning rate of $.0001$ and a batch size of $16$. Furthermore, we apply the dot product as compatibility function for our ranking loss explained in \Cref{ssec:zero-shot}. These settings and parameters were determined by preliminary experiments. The model state which performs best on the development set is then employed for the evaluation on the test set. The best model state is selected based on its macro F1-score which is a balance between precision and recall. The results and the corresponding discussion are presented in \Cref{sec:results}. 

\section{Results}
\label{sec:results}

\begin{table}[t]
  \centering
  \resizebox{.8\columnwidth}{!}{%
  \begin{tabular}{llrrrrrr}
  	\toprule
    \multicolumn{1}{l}{} & \multicolumn{3}{c}{\textbf{Dev}} & \multicolumn{3}{c}{\textbf{Test}}\\
    \cmidrule(lr){2-4} \cmidrule(lr){5-7}
    \multicolumn{1}{l}{\textbf{Embeddings}} & \textbf{ACC} & \textbf{UAR} & \textbf{F1} & \textbf{ACC} & \textbf{UAR} & \textbf{F1}\\
    \midrule
    \textsc{Bert} & .220 & .195 & .169 & .188 & .208 & .167\\
    \midrule
    \textsc{Avonet} & .372 & \textbf{.298} & \textit{.262} & .267 & .215 & .191\\
    \midrule
    \textsc{BLH} & \textbf{.384} & \textit{.288} & \textbf{.265} & \textbf{.289} & \textit{.286} & \textit{.221}\\
    \midrule
    \textsc{SBert} & .306 & .238 & .219 & .197 & .185 & .163\\
    \midrule
    \textsc{Bert}+\textsc{Avonet}+\textsc{BLH} & .181 & .175 & .154 & .175 & .168 & .151\\
    \midrule
    \textsc{Bert}+\textsc{Avonet} & .254 & .193 & .178 & .169 & .158 & .141\\
    \midrule
    \textsc{Bert}+\textsc{BLH} & .198 & .183 & .164 & .164 & .178 & .141\\
    \midrule
    \textsc{Avonet}+\textsc{BLH} & \textit{.335} & .281 & .244 & \textit{.287} & \textbf{.295} & \textbf{.233}\\
    \bottomrule
  \end{tabular}
  }
  \caption{The mean results over the development (Dev) and test (Test) sets of the five splits from \Cref{ssec:experimental-setup}. The best performance of each metric is marked bold, the second best is marked italic. The displayed metrics are \ac{ACC}, \ac{UAR}, and unweighted F1-score (F1). The F1-score poses the main evaluation metric.}
  \label{tab:results-linear-cv5}
\end{table}

For each test set of the five splits, we have $8$ to $10$ species which implies a chance \ac{UAR} between $10$\,\% and $12.5$\,\% to pick the correct class. Regarding the development sets we have $9$ to $11$ species entailing a chance level between $9.1$\,\% to $11.1$\,\%. The results of our experiments are listed in \Cref{tab:results-linear-cv5} and show the mean performance over all five splits. Since our experiments were optimised w.\,r.\,t.\ the F1-score, this is also the decisive metric regarding their evaluation. 
Therefore, the best-performing setting is the concatenation of the \textsc{Avonet} and \textsc{BLH} features. This outcome makes especially sense considering that the \textsc{BLH} features achieved the second and the \textsc{Avonet} the third best performance. 

Interestingly, the bird sound descriptions which we encoded with \textsc{Bert} and \textsc{SBert} perform worse than the morphological, ecological, and life-historical meta features. \textsc{Avonet} and \textsc{BLH} even outperform both encodings on all three tabulated metrics of \Cref{tab:results-linear-cv5}. This may be because the pre-trained language models have likely never seen bird-specific onomatopoeia such as ``vist, tk-tk-tk'' or ``si-sr\"u TILL-ILL-ILL-ILL-ILL....... (krschkrschkrsch) SR\"Usvisvi'' in the description example quoted in \Cref{sec:dataset} and thus might omit this information completely. 

Since both language model embeddings performed on an equal level, we only investigated the concatenation of the \textsc{Bert} embeddings with the other meta feature sets. Unlike the concatenation of \textsc{Avonet} with \textsc{BLH}, the concatenation of \textsc{Bert} with the functional and life-historical feature sets does not lead to an improvement, but rather a deterioration. 

As \Cref{fig:heatmaps} from \Cref{ssec:features} suggests that \textsc{SBert} should create better distinguishable embeddings than simple \textsc{Bert}, we furthermore expect to see a noticeable difference in their performances. However, \textsc{SBert} even performs slightly worse than \textsc{Bert}. This might be due to \textsc{SBert} putting more focus on semantic meaning than \textsc{Bert} which is difficult to achieve when the onomatopoeia are not properly considered and thus, crucial information is discarded. The difference becomes more obvious when consulting the results of \Cref{tab:results-linear-cv5}. The mean F1-score over the development and the test sets is nearly the same for \textsc{Bert} while there is a high discrepancy for \textsc{SBert} features, which may be a sign of overfitting.


\section{Conclusion}
\label{sec:conclusion}

We investigated three different sources of meta information for zero-shot audio classification of bird species: \textsc{(S)Bert} embeddings of textual descriptions of bird sounds, \textsc{Avonet} features comprising
functional traits
, and \textsc{BLH} features containing bird life-historical characteristics.
Our results suggest that the concatenation of \textsc{Avonet} and \textsc{BLH} achieve the best performance with a mean F1-score of $.233$ over five disjoint test sets, followed by solely utilising the \textsc{BLH} and \textsc{Avonet} features with a mean F1-score of $.221$ and $.191$, respectively. Therefore, the morphological, ecological, and life-historical meta information outperformed the encoded bird sound descriptions. We hypothesise that this is due to the language models not being pre-trained or fine-tuned on bird-specific onomatopoeic words or sentences. 

Future work could be to pre-train or fine-tune existing language models in order to better deal with onomatopoeic words and sentences so that this information is included in the embeddings. 
Furthermore, images of the bird species should be considered as another auxiliary information which can be properly encoded and processed together with the audio samples.
Since our goal was not to achieve state-of-the-art performance but to investigate different meta-features for our task, next steps could also be to improve the general model performance by means of employing and adapting more sophisticated \ac{ZSL} models.

\section{Acknowledgement}
\label{sec:acknowledgement}
This work was funded from the DFG project No.\ 512414116 (HearTheSpecies) and the DFG project No.\ 442218748 (AUDI0NOMOUS).




\vfill\pagebreak


\section{References}
\printbibliography[heading=none]

\end{document}